\documentclass[conference]{IEEEtran}
\IEEEoverridecommandlockouts

\usepackage{amsmath,amssymb,amsfonts, amsthm}
\usepackage{algorithmic}
\usepackage{booktabs}
\usepackage{cite}
\usepackage{comment}
\usepackage{courier}  
\usepackage{enumerate}
\usepackage{graphicx}
\usepackage{hyperref}
\usepackage{ifthen}
\usepackage{listings}
\usepackage{makecell}
\usepackage[charter]{mathdesign}
\usepackage{multirow}
\usepackage{pifont}  
\usepackage{stfloats} 
\usepackage{textcomp}
\usepackage[flushleft]{threeparttable}
\usepackage{url}
\usepackage{varwidth}
\usepackage{verbatim}
\usepackage{xcolor}

\def\BibTeX{{\rm B\kern-.05em{\sc i\kern-.025em b}\kern-.08em
    T\kern-.1667em\lower.7ex\hbox{E}\kern-.125emX}}

\definecolor{linkcolor}{HTML}{3bba3b}
\hypersetup{
	unicode = true,
	colorlinks = true,
	citecolor = linkcolor,
	filecolor = linkcolor,
	linkcolor = linkcolor,
	urlcolor = blue,
	pdfstartview = {FitH},
	breaklinks = true,
}

\begin{document}

\title{PQFabric: A Permissioned Blockchain Secure from Both Classical and Quantum Attacks
}
\author{
\IEEEauthorblockN{Amelia Holcomb}
\IEEEauthorblockA{\textit{Computer Science} \\
\textit{University of Waterloo}\\
Waterloo, Canada \\
aholcomb@uwaterloo.ca}
\and
\IEEEauthorblockN{Geovandro Pereira\thanks{G. Pereira and M. Mosca are supported in part by NSERC, CryptoWorks21, Canada First Research Excellence Fund, Public Works and Government Services Canada, and the Royal Bank of Canada.}}
\IEEEauthorblockA{\textit{Combinatorics \& Optimization} \\
\textit{University of Waterloo}\\
Waterloo, Canada \\
geovandro.pereira@uwaterloo.ca}
\and
\IEEEauthorblockN{Bhargav Das$^\ast $\thanks{$^\ast$ B. Das is currently affiliated with Indian Institute of Technology Dhanbad.}}
\IEEEauthorblockA{\textit{Computer Science} \\
\textit{University of Waterloo}\\
Waterloo, Canada \\
bhargav19036@gmail.com}
\and
\IEEEauthorblockN{Michele Mosca}
\IEEEauthorblockA{\textit{Combinatorics \& Optimization} \\
\textit{University of Waterloo}\\
Waterloo, Canada \\
michele.mosca@uwaterloo.ca}
}

\maketitle

\begin{abstract}
Hyperledger Fabric is a prominent and flexible solution for building permissioned distributed ledger platforms. Access control and identity management relies on a Membership Service Provider (MSP) whose cryptographic interface only handles standard PKI methods for authentication: RSA and ECDSA classical signatures. Also, MSP-issued credentials may use only one signature scheme, tying the credential-related functions to classical single-signature primitives. RSA and ECDSA are vulnerable to quantum attacks, with an ongoing post-quantum standardization process to identify quantum-safe drop-in replacements. In this paper, we propose a redesign of Fabric's credential-management procedures and related specifications in order to incorporate hybrid digital signatures, protecting against both classical and quantum attacks using one classical and one quantum-safe signature. We create PQFabric, an implementation of Fabric with hybrid signatures that integrates with the Open Quantum Safe (OQS) library. Our implementation offers complete crypto-agility, with the ability to perform live migration to a hybrid quantum-safe blockchain and select any existing OQS signature algorithm for each node. We perform comparative benchmarks of PQFabric with each of the NIST candidates and alternates, revealing that long public keys and signatures lead to an increase in hashing time that is sometimes comparable to the time spent signing or verifying messages itself. This is a new and potentially significant issue in the migration of blockchains to post-quantum signatures. 
\end{abstract}


\section{Introduction} \label{sec:introduction}
In recent years, the research community has drawn ever closer to the construction of a quantum computer, one with the potential to break classical public key cryptography. In advance of this threat, the National Institute of Standards and Technology (NIST) is looking to establish standards for cryptographic algorithms that show promise at securing against both quantum and classical attacks. The NIST post-quantum standardization process is in its third round of evaluation and has narrowed down the initial 82 submission candidates to only 15. These include six digital signature candidates: three finalists and three alternates. Some of the finalists are expected to be selected for standardization at the end of Round 3, about an year from now. A fourth round is also expected in order to select alternate candidates for later standardization.

Open source implementations of such algorithms, for instance those provided by the Open Quantum Safe (OQS) project \cite{oqsproject}, continually change in order to keep up-to-date with both progress in cryptanalysis and per-algorithm parameter modifications and optimizations. However, designing, approving, and implementing post-quantum cryptographic standards is only the first step. There is also the gargantuan task of integrating these algorithms into the wide range of existing systems that depend on and are highly coupled with classical public key cryptography. The integration itself will be a new test of these algorithms, particularly as they are integrated into high-performance systems. Moreover, during the transition, these systems must remain secure against classical attacks, even in the face of potential flaws or changes to post-quantum cryptography algorithms. This situation demands a double layer of protection against classical and quantum attacks, which can be achieved by means of hybrid signatures, i.e., one classical signature and one post-quantum \cite{bindel2017transitioning,crockett2019prototyping}. 

A prominent class of applications that is likely to be an early adopter of these algorithms is the blockchain distributed ledger. Blockchain technology is fundamentally reliant on cryptographic primitives, especially signature authentication, in order to 1) validate the origin of the transactions that are to be added to the chain and 2) ensure non-repudiation from the transaction originators. In our work, we focus on Hyperledger Fabric \cite{androulaki2018fabric}, a permissioned blockchain in use in production systems. As a permissioned blockchain, Hyperledger does not require anonymous credentials, but instead uses standard digital signatures to identify users.

In addition to the need for blockchains to make this post-quantum transition earlier rather than later, there are several typical features of blockchains that make them a rigorous testbed for new cryptographic algorithms. Industry-deployed blockchains are high-throughput, with recent research in Hyperledger Fabric achieving 20,000 transactions per second \cite{gorenflo_fastfabric_2020}. Each transaction proposal along with every processing step in the network contains a signature, and the code path to commit each transaction includes both signing and verification algorithms. As such, signature size and algorithm speed both have direct and significant impact on the performance of the system. In addition, as large distributed systems that can neither ensure synchronous updates nor handle extended downtime for a migration, blockchains are among the more complex use cases for post-quantum migration. Lessons learned maintaining performance and backwards compatibility in this context can be broadly applicable as other systems integrate with post-quantum algorithms later on. 

\paragraph*{Contributions} This work introduces PQFabric, which is to our knowledge the first version of the Hyperledger Fabric enterprise permissioned blockchain\footnote{the FastFabric project \cite{gorenflo_fastfabric_2020} in particular.} whose signatures are secure against both classical and quantum computing threats. 
Our design proposal is backwards compatible for easy migration and flexible enough to support different post-quantum cryptographic schemes, making it future-proof to NIST upcoming standards and updates to the OQS library. We also: \begin{itemize}
    \item Highlight the inconsistencies we found between NIST and Golang APIs if hybrid signature schemes are to be implemented, which will need to be resolved for smooth integration going forward.
    \item Provide a baseline performance comparison between transactions that use classical and hybrid quantum-classical cryptography. We find that, depending on the signature algorithm used, PQFabric can achieve comparable performance to classical Fabric.
    \item Profile PQFabric to provide insight into the practical performance costs of post-quantum signature algorithms. We show that the length of the signature and/or public key, not just the signature algorithm execution time, are major contributors to PQFabric slowdowns.
    \item Publish our code online, offering our implementation as a testbed for future work focusing on improving blockchain transaction throughput with post-quantum cryptography algorithms.
\end{itemize}

\section{Review of Hyperledger Fabric and Related Work}\label{sec:background}
Hyperledger Fabric \cite{androulaki2018fabric} consists of a highly flexible and customizable framework for deploying permissioned blockchains. 
Fabric achieves high performance and scalability through the execute-order-validate paradigm, which was originally proposed in the context of improving the performance of State Machine Replication \cite{kapritsos2012}. This means that particular peers execute the transactions first for prevalidation before they transactions get ordered. This is in contrast with the typical approach from other blockchains of ordering the transactions first and then requiring all peers to execute all ordered transactions (order-execute-validate). This modified approach allows for a number of optimizations like avoiding all peers to execute every transaction among many others. Moreover, operations like execution/validation and ordering of the transactions are decoupled and thus performed by different entities (peers/endorsers and orderers, respectively). As a result, the consensus protocol (run by the orderers) is independent of the blockchain protocol as a whole (e.g., the chaincode), and can be tailored per application and also updated after the network is bootstrapped. 

We now provide a set of definitions of interest used in the context of the Fabric Blockchain.

\textbf{Membership Service Provider (MSP).} The network MSP handles credential/identity management and is usually associated with one organization. It is responsible for issuing certificates to peers, orderers and users and thus giving them suitable authentication and authorization. The network MSP is not to be confused with the local MSP, which is discussed in Section \ref{sec:implementation}. Each node uses a local MSP to abstract credentials and identity management (including signatures and their verification) from the rest of its codebase. The local MSP is a software module only, and does not run separately from the node.

\textbf{Fabric-CA}. The Fabric-CA is a certificate authority that contains a root certificate and issues identity certificates to each organization's MSP. The MSP's certificate allows it to issue enrollment certificates to the users in its organization, giving them access to the blockchain. 

\textbf{Client or User.} A client or user is an actor who owns digital access credentials to a particular blockchain network, and can submit transaction proposals to the peers.   

\textbf{Peer (also called endorsing peer or endorser)}. The peer's responsibility includes verifying users' signatures in the transaction proposals, executing and validating the proposal according to pre-specified policies, endorsing and digitally signing the resulting outcome of the validated proposal, and notifying the user of the outcome. Peers also receive block (of ordered transactions) candidates from the orderers, performing final validation and committing the blocks to the ledger.

\textbf{Orderer.} For each set of transactions, an orderer verifies all endorsers' signatures and runs a consensus protocol to order the transactions into a block candidate. Orderers sign the block candidate and send it back to peers for final validation and inclusion in the ledger.


\textbf{Transaction.} Users submit transaction proposals to the blockchain network in order to change the state of  the ledger. A proposal includes the user's identity, the chaincode (smart contract) function to be executed and its related parameters, and a transaction identifier. The user is responsible for soliciting endorsements according to the organization's policies. The user assembles the proposal and endorsements, signs everything, and submits it to the ordering service. The submitted content is the transaction.


\subsection{Related Work}
We now briefly review other blockchain proposals in the literature aiming for quantum-safe solutions.
We recall that the goal of this work is to provide the main Public Key Infrastructure (PKI) functionalities with quantum-safe protections. PKI in Fabric allows for reliable authentication and authorization throughout the blockchain interactions. Protecting the PKI consists of migrating classical digital signature primitives into hybrid ones that contain a post-quantum counterpart. Much of the related work in the literature does not target the same goal. 

A common quantum-safe proposal in the literature is to protect the classical Confidential Transactions protocol also named RingCT. The RingCT protocol was tailored for use in cryptocurrency applications. Its main properties include anonymity of the user executing transactions, concealed transaction amounts, and provable ranges for the transaction amounts. This is the case in proposals like Lattice RingCT \cite{sun2017ringct,torres2018post}, Hcash \cite{hcash} and MatRiCT \cite{esgin2019matrict}. The goal in such proposals is to provide quantum-safe protection for anonymous clients running on a public blockchain, i.e., clients who do not want to have their identity revealed when performing transactions. Anonymous credentials can be achieved by the means of ring signatures \cite{rivest2001leak}. NIST is far from standardizing post-quantum ring signatures as we are still in the process of selecting a plain digital signature in the next few years. Hyperledger Fabric, as a permissioned blockchain, does not adopt anonymous credentials.

In 2019, Campbell \cite{campbell_transitioning_2019} evaluated the use of qTesla, one of NIST's Round 2 post-quantum signature candidates, in Hyperledger Fabric. Although the paper discusses the principle of hybrid signatures, it ultimately examines only a fixed non-hybrid signature scheme (qTesla), while we implement and analyze hybrid signatures that are configurable with any post-quantum signature algorithm. Moreover, its analysis uses only the published benchmarks of qTesla, and provides no practical impact evaluation of integrating hybrid post-quantum signatures into Fabric nor design recommendations for the more subtle elements of the integration.


\section{Solution Proposal}\label{sec:proposal}
\subsection{Signature Requirements}\label{Representation}
When designing a hybrid quantum-classical signature for use in Hyperledger Fabric, we have two key requirements. 
\begin{enumerate}
    \item In the future quantum computing world, the system must be as protected as the strongest currently developed post-quantum cryptography standards will allow.
    \item In the current classical computing world, the system must be no less protected than current cryptography standards.
\end{enumerate}
To satisfy the first requirement, it would be straightforward to use a post-quantum algorithm directly. Previous work \cite{campbell_transitioning_2019} deemed the qTesla signature algorithm \cite{noauthor_qtesla_nodate} alone sufficient to meet both requirements, because all qTesla parameters were at the time believed to be theoretically secure against both quantum and classical attacks. However, we argue that this is not sufficient, even if qTesla was theoretically secure. A cryptographic algorithm is only as secure as its implementation, and these new implementations are relatively untested and unproven. If a flaw is found in the implementation of a post-quantum algorithm, we must still have the classical signature. Otherwise, a revealed vulnerability in the post-quantum implementation would violate our second requirement.

\subsection{System Requirements}
In designing PQFabric, we have two system requirements.
\begin{enumerate}
    \item \textit{Backwards compatibility:} Fabric is a large distributed system whose nodes and clients cannot upgrade simultaneously. In addition, we do not expect that organizations running fabric will retire their existing blockchain and start a new one from a genesis block in order to migrate to post-quantum cryptography. As a result, our solution must be backwards-compatible. The blockchain must be able to gradually migrate from classical cryptography to post-quantum cryptography without unreasonable or synchronous downtime, and client software running classical cryptography must be able to coexist with PQFabric.
    \item \textit{Cryptoagility:} Our solution must be flexible to ongoing changes in post-quantum cryptographic standards, as NIST continues to finalize the list of post-quantum signature algorithms. It must be compatible with all the remaining candidates and agnostic to which algorithm is eventually chosen. Selecting any of them must be a straightforward configuration change. 
\end{enumerate}
\subsection{Transaction Signature Proposal}
We propose a hybrid quantum-classical signature scheme for transactions, using both post-quantum and classical cryptography. Each node in Fabric has two private keys, one classical and one post-quantum. Nodes sign transactions once with each key and concatenate the two signatures. Both signatures must be verified.

\subsection{Identity Proposal} \label{sec:proposal:identity}
Hyperledger Fabric passes public key identities as X.509 certificates. We follow the proposal in \cite{kampanakis_viability_2018} to create hybrid X.509 certificates, adding three non-critical Extensions:
\begin{itemize}
    \item Subject-Alt-Public-Key-Info: The post-quantum public key for the certificate, which may be nil.
    \item Alt-Signature-Value: The signed classical and post-quantum public key material of the certificate, signed by the issuer's post-quantum key.
    \item Alt-Signature-Algorithm: The post-quantum algorithm used to sign the certificate key material.
\end{itemize}
The issuer still signs the entire certificate with a classical-only signature. Because the extensions containing the post-quantum material can be ignored by classical verifiers, the hybrid certificates are backwards compatible. 

\subsection{Security analysis of the proposal}

\subsubsection{Hybrid Signatures} There are different combiners for building hybrid post-quantum + classical signatures, whose security is generally analyzed under the Existential Unforgeability under Chosen Message Attack (EUF-CMA). We consider the EUF-CMA security of both the certificate signatures and the transaction signatures.

A dual message combiner with nesting is suitable for dual messages $(m_1, m_2)$ like those in hybrid X.509 certificates\footnote{$m_1$ is a non-critical X.509 Extension}. In particular, the dual combiner signs as follows: 
\begin{align*}
    \sigma_1 &\gets \Sigma_1.\textit{Sign}(m_1) \\
    \sigma_2 &\gets \Sigma_2.\textit{Sign}(m_1,\sigma_1,m_2) \\
    \sigma &\gets (\sigma_1, \sigma_2)
\end{align*}
where $\Sigma_i$ refers to the respective classical or post-quantum algorithm, and the combined signature is $\sigma$. According to \cite{bindel2017transitioning}, if $\Sigma_1$ is EUF-CMA secure then the combined signature is EUF-CMA secure only with respect to its first component $\sigma_1$. Moreover, if $\Sigma_2$ is EUF-CMA secure then the overall $\sigma$ is also EUF-CMA-secure, i.e., the outer signature  guarantees the unforgeability of both messages $(m_1,m_2)$. We follow the above construction for certificate signatures. $\Sigma_1$ is a post-quantum algorithm; $\Sigma_2$ is a classical algorithm; $m_1$ is the post-quantum public key material; $m_2$ is the classical certificate information that is signed classically by the issuer along with Extensions containing $m_1$ and $\sigma_1$. 

The dual message combiner is backwards-compatible with legacy signers and verifiers. If a node cannot use the post-quantum layer, $\sigma_1,m_1 \gets Null$ reverts to classical-only authentication. This approach also resists downgrade (to classical-only) attacks, since an attacker cannot detach the pure classical signature component from $\sigma$ as long as the classical signature is still secure. Formally, they cannot retrieve $\sigma_2' = \Sigma_2.Sign(m_2)$ from the dual signature $\sigma_2 = \Sigma_2.Sign(m_1, \sigma_1, m_2)$. Note the separation would be easy if a concatenation combiner\footnote{$\sigma = (\Sigma_1.Sign(m_1) || \Sigma_2.Sign(m_2))$} was used.

Next, we consider the security of the transaction signature. The EUF-CMA security of a combiner is given by the maximum EUF-CMA security offered between the two signatures \cite{bindel2017transitioning} if single messages are to be signed. In the case of the hybrid signatures generated by Fabric peers, orderers, and clients, a concatenation combiner is sufficient. Fabric uses provided identities, in the form of X.509 certificates, to verify signatures, and stores these identities with each signature in committed blocks. Keeping with the conventions of existing Fabric, we use the X.509 certificate, not the signature itself, to decide whether the signature should be interpreted as hybrid or classical-only. Thus, downgrade attacks on the concatenated signature do not apply, because it is protected by a nested signature (the certificate). 


We also emphasize that nested signatures at the transaction level would add significantly to transaction times, because they require hashing potentially huge messages (recall that the post-quantum signature $\sigma_1$ is appended to the message). For example, Picnic signatures are 34--61KiB.

\subsubsection{Hashing} Hashing was not the primary focus of our work; however it does have implications for post-quantum security. Cryptographic hashes are not broken by quantum computers, but their preimage-resistance security is reduced by half and collision-resistance is reduced from $N/2$-bit security to about $N/3$ bits for an $N$-bit output hash \cite{grover1996fast,bennett1997strengths}. For this reason, $N \ge 384$ is recommended for collision-resistant post-quantum hashing. We upgraded signature hashes to SHA-384 in PQFabric, but Fabric does not yet offer pluggable hash functions for the block hashes forming the ledger. This is under development by the Hyperledger team.

\section{Implementation}\label{sec:implementation}
In a 2018 guide for businesses using Hyperledger Fabric, IBM researchers wrote in a brief section on quantum cryptography that “Hyperledger Fabric provides a pluggable cryptographic provider, which allows replacing these [existing] algorithms for digital signatures with others.” \cite{gaur_hands-blockchain_2018} However, the process is not that simple. In this section, we will describe the basic structure of our implementation followed by the challenges we faced implementing hybrid quantum-classical scheme.

\subsection{Core Structure}
In our implementation,\footnote{Code at \url{https://github.com/ameliaholcomb/fastfabric1.4-oqs}} we built off of Hyperledger Fabric 1.4. We used LibOQS 0.4.0 \cite{stebila_post-quantum_2017} for the implementations of post-quantum cryptographic signature algorithms, which was the most recent version at the time of writing, but our code is compatible with all versions of LibOQS. LibOQS is written in C while Hyperledger Fabric is written in Go, so we wrote a CGO wrapper around LibOQS.\footnote{Recently, libOQS has published its own Go wrapper, which can be found at \url{https://github.com/open-quantum-safe/liboqs-go}.} Eventually, we assume that these quantum-safe cryptographic functions would be built into the Go core library, so we followed the API conventions used in Go's ECDSA implementation.



We introduced a \texttt{SecretKey}, a \texttt{PublicKey}, and an \texttt{OQSSigInfo} struct attached to each containing the OQS algorithm name.
The algorithm may be set by configuration and is not expected to change over the lifetime of the blockchain except by careful migration (see Appendix \ref{sec:appendix:migration}).\footnote{That said, any node \textit{can} verify any post-quantum signature for which it has the appropriate algorithm implementation, even if it is not configured to sign with that algorithm itself. } The Go representation of the LibOQS library and \texttt{Sig} objects are loaded together and maintain pointers to LibOQS C functions. In our implementation, post-quantum cryptography is only used for signing and verifying messages, so the wrapper only includes \texttt{KeyPair}, \texttt{Sign}, and \texttt{Verify}. 

We modified three main areas of the Hyperledger Fabric codebase to allow hybrid quantum signatures. \begin{enumerate}
    \item Blockchain Cryptographic Service Provider (BCCSP): The BCCSP module is designed to provide a uniform cryptography interface to the core Fabric that is not dependent on a particular cryptographic algorithm or implementation. It is a specific implementation of the more general Cryptographic Service Provider (CSP). The bulk of our modifications here were to add a new key type and associated interface with \texttt{KeyImport}, \texttt{KeyPair}, \texttt{Sign}, \texttt{Verify}, and so on. We also made changes to the \texttt{Signer} module shared by all BCCSP keys; for more details, see section \ref{sec:signer}.
    \item Local Membership Service Provider (MSP): In theory, the BCCSP modifications should be all that is necessary. However, as discussed in section \ref{sec:challenges:twokeys}, hybrid quantum-classical cryptography requires two keys, which is not a transparent change for the MSP module.\footnote{The MSP also needed to be modified because, while signing functionality is fully factored out into a shared \texttt{Signer} in the BCCSP module, the corresponding verify functions are not. Presumably this is an oversight; it would not be difficult to add a \texttt{Verifier} in the BCCSP.}
    \item Cryptogen: This binary is not an officially supported part of Hyperledger Fabric, but it provides a template for generating the cryptographic material required to run Fabric from its configuration files. Organizations running Fabric may generate this material however they choose, but any equivalent of the ``cryptogen'' binary would need similar modification in order to generate post-quantum key material and X.509 certificates. 
\end{enumerate}

One might also modify the client to use quantum-safe cryptography. Since we were primarily concerned with the integrity of the blockchain as a whole, rather than transactions submitted by a single client, we did not include this in the scope of our work. Moreover, as the client code is written in Javascript and makes no use of the MSP or BCCSP common packages, it is a large disjoint refactoring project with limited interest to the research community. 

\subsection{Signature Structure}\label{sec:signer}

The Go crypto library Signer interface \cite{noauthor_crypto_nodate} has methods
\begin{lstlisting}
Public() PublicKey
Sign(rand io.Reader, digest []byte, opts SignerOpts) (signature []byte, err error)
\end{lstlisting}

Hyperledger Fabric's BCCSP module offers an implementation of the above \texttt{crypto.Signer} interface that calls the appropriate signature algorithm based on the key type used in Signer creation. We modified the BCCSP Signer to implement hybrid signatures.

We added an optional post-quantum key field to the \texttt{bccsp.Signer} struct, filled out at Signer creation. The very careful reader may notice that this modification will not quite match the interface of Go's \texttt{crypto.Signer}. First, the crypto library explicitly assumes that there is only a single public key, and enforces this assumption in the type signature of the \texttt{Public()} method. We discuss this problem, as it appeared here and elsewhere, in \ref{sec:challenges:twokeys}. Second, there is a subtle incompatibility between the crypto library and LibOQS: while \texttt{crypto.Signer.Sign()} requires a digest, LibOQS's \texttt{Sign()} functions expect an unhashed message. We discuss this issue further in \ref{sec:challenges:hashing}.

Setting aside these discrepancies, we describe the hybrid signature scheme used. We use concatenation as follows:
\begin{lstlisting}
if qKey {
    s1 = Sign(qKey, digest)
}
s2 = Sign(cKey, digest)
if qKey {
    return asn1.Marshal(s1, s2)
}
return s2
\end{lstlisting}

We make the corresponding modifications to verification. We first parse and verify the provided X.509 identity, checking the alternate public key extensions to decide whether or not the identity is post-quantum. If the X.509 certificate contains an alternate public key, we unmarshal the signature and verify the quantum and classical parts separately. Both must match for the signature to be accepted. If there is a post-quantum certifying authority, the X.509 certificate must contain post-quantum extensions even if the client is classical-only, with a quantum signature certifying the absence of a post-quantum public key. In this case, the client is considered ``legacy'' and the node accepts a classical-only signature. If the certifying authority has only a classical key, then PQFabric behaves just like Fabric.

\subsection{Integration Challenges}\label{sec:challenges}

Overall, the main Hyperledger Fabric codebase did a remarkably good job of encapsulating its cryptographic interface in the BCCSP and making few assumptions about key or algorithm type. We suspect integrating other codebases with quantum-safe cryptography may be more of a refactoring effort. (For example, moving outside of the well-structured Fabric code to the auxiliary Cryptogen binary code, we quickly encountered functions like \texttt{GetECPublicKey}, which can return only an ECDSA public key.) However, there were a few challenges in the integration that may help reveal the complex code assumptions that will change under a hybrid quantum-classical cryptography scheme.

\subsubsection{Two Keys}\label{sec:challenges:twokeys}
The hybrid quantum-classical cryptography scheme necessitated by the requirements in \ref{Representation} implies that two private (and public) keys are needed. One might envision creating a HybridKey type that encapsulates two lower-level keys. However, \texttt{Sign()} is a method on a Signer or other higher level object, and hybridization is fundamentally a change to signatures, not to keys. Thus, we implemented a hybrid Signer type instead. The hybrid Signer conflicted with assumptions at both the higher MSP layer and the lower Go crypto library layer.

First, placing an additional key in the Signer broke the encapsulation between the BCCSP and MSP modules, because the MSP:
\begin{itemize}
    \item Extracts cryptographic keys from an X.509 certificate. It does this both on initialization (extracting public keys from its own X.509 certificate and importing the corresponding private key from its keystore) and during signature verification (deserializing the public keys from another node's identity proto).
    \item Stores the public and private keys for the node directly in an internal proto structure called an \texttt{Identity}.
    \item Provides its own stored keys to the BCCSP when signing a message.
\end{itemize}
In these three areas, the MSP had to be modified to extract, store, and use a second key.

Second, placing an additional key in the Signer conflicted with the Golang library because its \texttt{crypto.Signer} has an interface method \texttt{Public()}, which returns a single public key. In the case of Hyperledger Fabric, the \texttt{Public()} function is never used, and so we were able to sidestep the issue entirely. However, it is important to highlight that the Golang \texttt{crypto.Signer} interface is not currently compatible with hybrid Signers. 

\subsubsection{Hashing}\label{sec:challenges:hashing} 

Our PQFabric integration also revealed inconsistencies between the NIST standard specifications and the core Golang crypto library. Specifically, the NIST Sign API takes an unhashed message \cite{computer_security_division_example_2017}, while the Go crypto library uses a Signer interface that expects the actual digest \cite{noauthor_crypto_nodate}. Unfortunately, these divergent API decisions are both used as features. Several finalist algorithms have taken advantage of NIST's API to select an appropriate hash function internally based on the security parameters specified, or to provide optimizations when the same hash is required twice. Hyperledger Fabric, meanwhile, makes use of the Go library's flexibility in their first stages of implementing  pluggable hash functions.

In our implementation, we simply allowed LibOQS to internally re-hash a hashed message. This may have a small performance cost but does not impact security (it provides as much security as the less secure hash). We could have removed the hashing step on the Hyperledger Fabric side, but it would have taken heavy contortions to do this while still maintaining code compatibility with a purely classical Fabric. Unless Golang and NIST's APIs agree, this is likely to be a problem for future integrations as well.

\subsubsection{Backwards Compatibility}\label{sec:challenges:backcompat}

Our scheme is backwards compatible in the sense that no individual node or client has to use post-quantum keys: every node checks whether the signing identity contains a post-quantum key before deciding whether a hybrid signature verification is required. This allows, for example, all the peers and orderers to use quantum-safe cryptography while the client still has not been upgraded (as we did). However, every node on the blockchain must at least have a software update that allows it to verify post-quantum hybrid signatures before any one node can use them. A vanilla fabric node will not know how to unpack or verify a hybrid signature, and we do not offer a classical-only component for partial verification. 

For an explicit proposal of how to perform a live migration to hybrid post-quantum cryptography, see Appendix \ref{sec:appendix:migration}.

\subsubsection{Cryptoagility}
In the existing code base, developers must add a new key type and implement its interface functions in the BCCSP package for each new algorithm. In order to use LibOQS algorithm implementations and keep up-to-date with changes to the NIST finalists at each stage, we needed to be algorithm-agnostic, without a separate key type for each post-quantum signature algorithm.

This presented some challenges for code integration because, unlike with other keys, the Go data type did not uniquely determine the algorithm to be used. We introduced the OQSSigInfo struct, a member of an OQS PublicKey, so that Go operations requiring an algorithm identifier, such as marshalling/unmarshalling keys or creating an Alt-Signature-Algorithm X.509 extension (see section \ref{sec:proposal:identity}), could obtain a key algorithm.

This challenge will not be an issue for developers who wait for complete standardization and a per-language implementation of post-quantum cryptography. However, in the meantime, while NIST solicits feedback on integration of these non-standard algorithms into production systems, maintaining cryptoagility is an additional consideration.

\section{Evaluation and Results}\label{sec:evaluation}
We evaluated our implementation on a network consisting of one orderer and two peers, each running on a different machine. The client (itself on a fourth machine) sent all transactions to a single peer, while the second peer was also available as an endorser. Each machine was equipped with Intel\textsuperscript\textregistered Xeon\textsuperscript\textregistered  CPU E5-2620 v2 processors at 2.1 GHz, with 24 hardware threads, 64GB RAM, and an SSD.

We use the chaincode from \cite{gorenflo_fastfabric_2020}, which provides simple balance accounts and allows transferring value from one account to another. A single transaction thus touches exactly two accounts. Our experimental setup includes 20,000 accounts, and each benchmark runs 10,000 transactions, batched into blocks of size 100, with 100 blocks sent from each of 10 parallel threads. The transactions were created to ensure that no two touched the same accounts, so that database contention would not be a factor. The endorsement policy required only one peer to endorse a transaction. The benchmark measured wall time on the peer between receiving a block of transactions and committing that block. For each round of benchmarks, we trimmed off the first and last few blocks in our analysis for ramp up and ramp down.

Our baseline cryptographic setup signs transactions only with ECDSA defined over the NIST curve P-256 (as specified by FIPS 186-3) and offers 128-bit classical security. We then compared this to the same benchmark run with nodes configured to sign and verify using hybrid schemes pairing ECDSA\footnote{Throughout the paper, we refer to the hybrid schemes by their post-quantum algorithm only.} with each remaining NIST round three finalist and some of the alternates, as implemented in libOQS 0.4.0. We used various post-quantum parameter sets targeting NIST security category I. Not all of these signature schemes were able to run successfully with Hyperledger Fabric. In particular, large certificate sizes caused various symptoms, including peer node crashes and stuckness in the endorser. We did not attempt to debug these issues, and instead report only on the performance of the post-quantum hybrids that ran successfully without modification (see Table \ref{tab:success}). 

Finally, though qTesla is no longer in the running for NIST, we still decided to evaluate its performance. We did this because it was specifically considered in recent work on post-quantum Hyperledger Fabric \cite{campbell_transitioning_2019}, and because it has interesting performance characteristics with a large certificate that is not quite of sufficient size to cause crashes.

\newcommand{\cmark}{\ding{51}}%
\newcommand{\xmark}{\ding{55}}%
\begin{table}
\caption{Tested algorithms, sorted by certificate size in bytes. Algorithms with too large a certificate caused the PQFabric peer to crash during the benchmark.}\label{tab:success}
\centering
\begin{tabular}{lrl}
\toprule
                    Algorithm &      Cert Size (bytes)    & Success \\
\midrule
                        ECDSA &                818 &            \cmark \\
                   Falcon-512 &              2,988 &            \cmark \\
                  Falcon-1024 &              5,051 &            \cmark \\
                  Dilithium-2 &              5,263 &            \cmark \\
                  Dilithium-3 &              6,542 &            \cmark \\
                  Dilithium-4 &              7,830 &            \cmark \\
                   qTesla-p-I &             24,551 &            \cmark \\
                 Picnic-L1-FS &             45,319 &            \xmark \\
 Rainbow-Ia-Cyclic-Compressed &             79,708 &            \xmark \\
           Rainbow-Ia-Classic &            202,741 &            \xmark \\
\bottomrule
\end{tabular}
\end{table}

\begin{figure}[!t]
      \centering
      \includegraphics[scale=0.32]{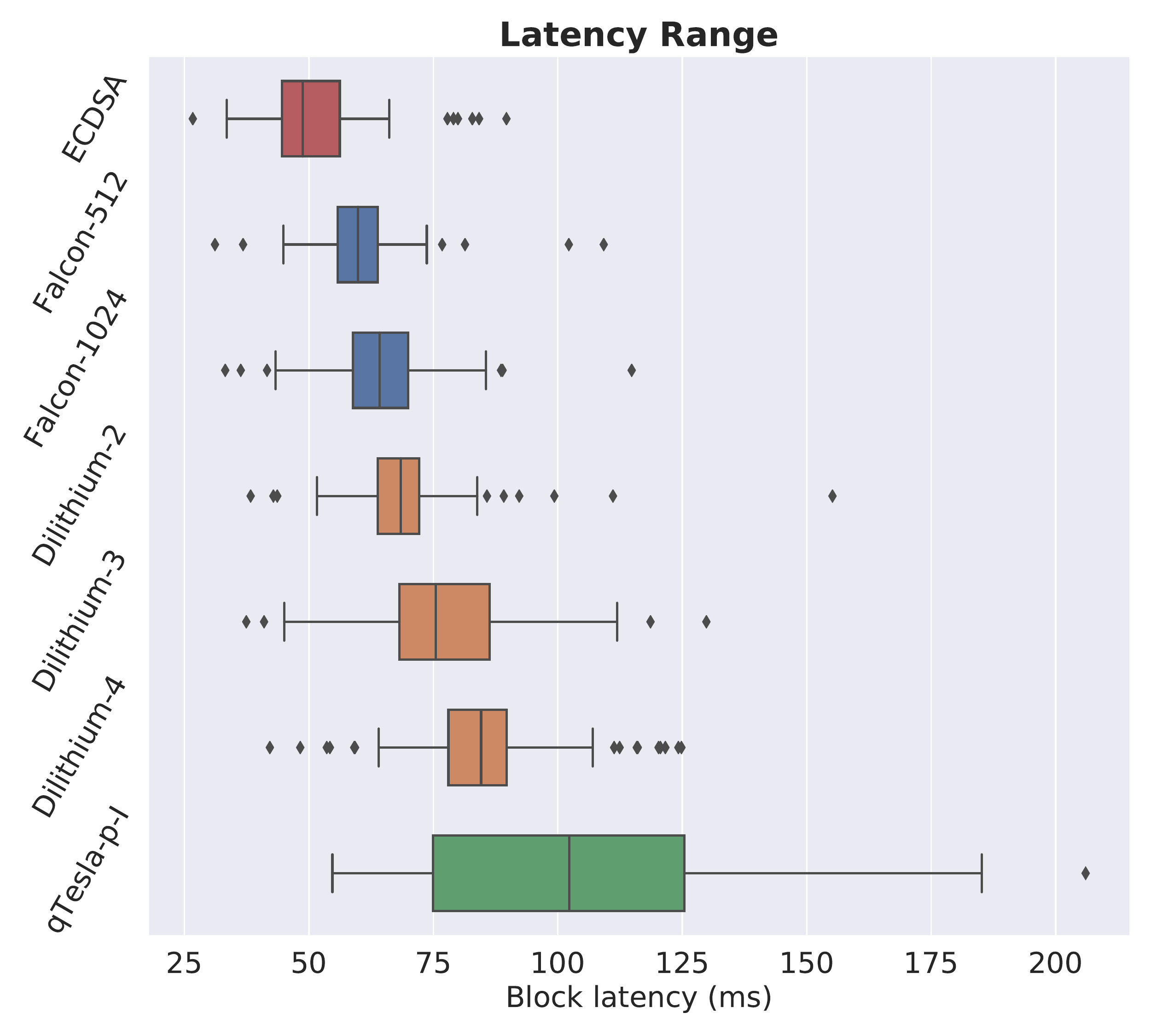}
      \caption{Per-block commit latency range for different signature algorithms and security parameters. Note that hybrid signatures are referred to by their post-quantum algorithm specification only.}
      \label{fig:latency}
      \vspace{-0.5cm}
\end{figure}
\begin{figure}[!t]
      \centering
      \includegraphics[scale=0.32]{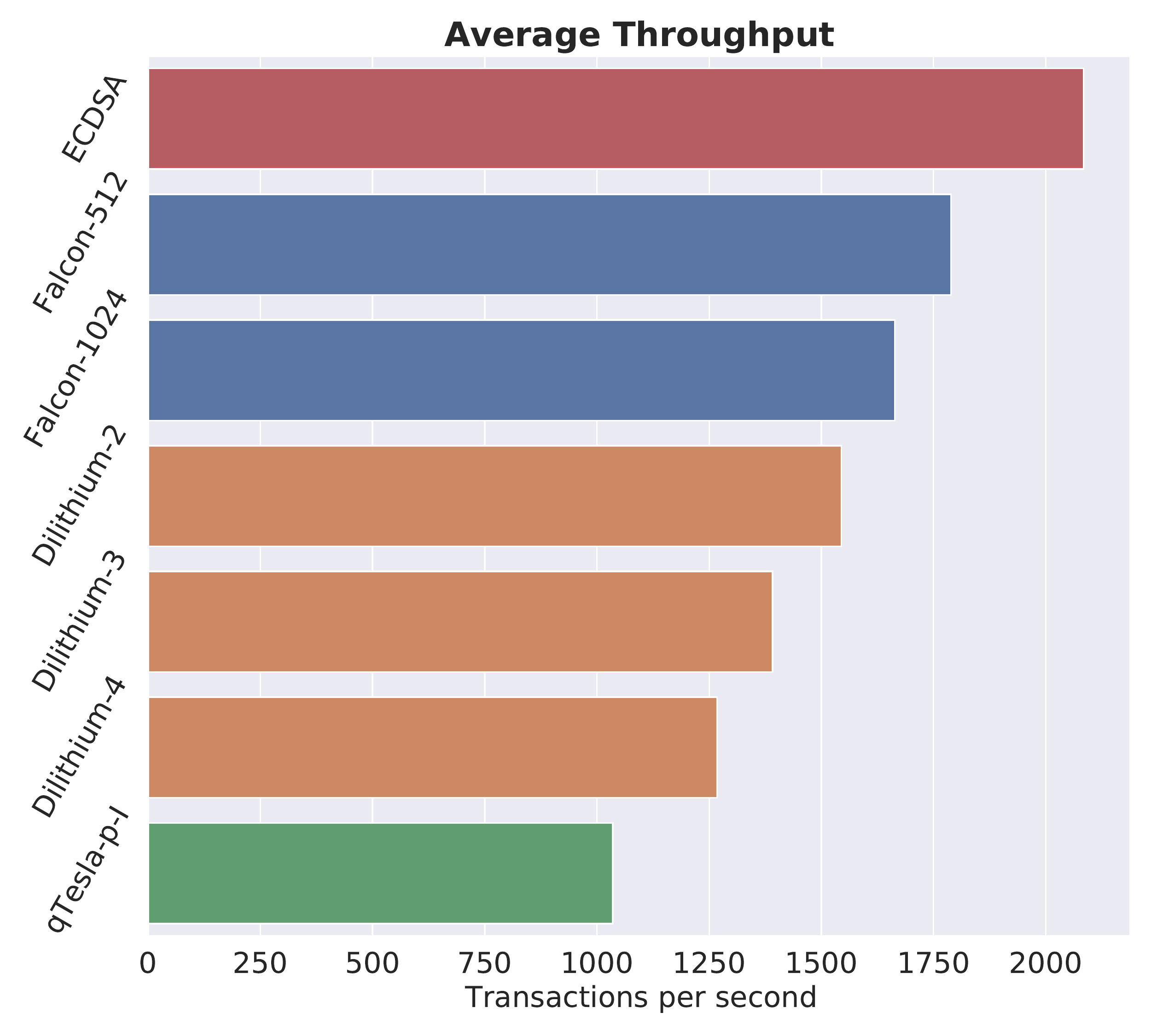}
      \caption{Average throughput, in transactions per second, of different signature algorithms and security parameters. Note that hybrid signatures are referred to by their post-quantum algorithm specification only.}
      \label{fig:throughput}
      \vspace{-0.5cm}
\end{figure}

As shown in Figure \ref{fig:latency}, ECDSA alone has the lowest block latency (median 49 ms), followed by the Falcon hybrids (median 60 ms and 64 ms), the Dilithium hybrids (median 68 ms, 76 ms, and 85 ms), and finally ECDSA+qTesla I (median 102 ms). Their average throughput, shown in Figure \ref{fig:throughput}, follows the opposite pattern, with throughputs of 2084, 1788, 1664, 1545, 1391, 1268, and 1035 transactions per second, respectively. We further investigated the source of the slowdown by examining representative CPU profiles of the peer node below.

\section{Discussion}\label{sec:discussion}
\subsection{Performance} \label{sec:discussion:performance}
Overall, we found that PQFabric, while having a higher block latency and lower throughput than classical Fabric, does not experience severe performance degradation, depending on the post-quantum signature algorithm used. The hybrid Falcon-512 scheme saw only a 14\% decrease in throughput, on average, compared to the pure ECDSA scheme. Indeed, the time spent in Falcon-512 Sign and Verify alone is faster than the corresponding ECDSA functions; much of the slowdown comes from having to sign and verify twice. This is significant and surprising: we expected hybrid signatures might severely slow down transactions, but a hybrid ECDSA+Falcon-512 scheme seems usable as-is with no further performance optimizations.

We now examine representative CPU profiles for benchmarks of each signature scheme. The functions with the greatest increase in CPU share for the hybrid schemes are \texttt{oqs.\{Sign$ \vert $Verify\}} and \texttt{sw.Hash}. The first is expected; this is the additional execution time required to perform the post-quantum signature algorithm, on top of the ECDSA algorithm. The time spent in Hash is more surprising. Some of the slowdown may be accounted to the SHA-384 hash required by post-quantum signatures, however that does not explain the variation between post-quantum algorithms. There are two main callers of Hash that contribute to its CPU time. The first is from signature verification in the MSP. The hash function is called on the received message. The second caller is VerifyBlock, usually from the gossip service. In this case, the hash function is called on the entire block. This block contains the identity certificates, including a public key and certificate signature, as well as message signatures for all of the endorsements received. For post-quantum signatures these may be quite large, dramatically increasing the CPU time spent on hashing. Table \ref{tab:hash} estimates the time each signature scheme spent on hashing and in LibOQS functions (Sign and Verify) for each block, compared to the public key plus signature size of the scheme.

\begin{table}
    \setlength{\tabcolsep}{3pt}
    \caption{Algorithm signature + public key size, versus benchmark data.}
    \label{tab:hash}
    \begin{tabular}{lrrrr}
\toprule
   Algorithm &  Pk+Sig Size & LibOQS Time$^{\mathrm{*}}$ & Hashing Time$^{\mathrm{*}}$ & Block Latency \\
             &      (bytes) &       (ms)  &  (ms)        & (ms) \\
\midrule
       ECDSA &           96 &          -- &            4 &            52 \\
  Falcon-512 &         1563 &          13 &            5 &            60 \\
 Falcon-1024 &         3073 &          22 &            6 &            65 \\
 Dilithium-2 &         3228 &           4 &           14 &            70 \\
 Dilithium-3 &         4173 &           5 &           16 &            77 \\
 Dilithium-4 &         5126 &           7 &           18 &            85 \\
  qTesla-p-I &        17472 &           8 &           29 &           104 \\
\bottomrule
\end{tabular}
    \begin{tablenotes}
      \item $^{\mathrm{*}}$ The LibOQS time and Hashing time should be taken as \textit{rough approximations only}. They were calculated by multiplying the mean block latency by the percent of time spent in LibOQS and Hash, as measured by a sample CPU profile of the entire benchmark. The sample profile includes setup blocks, warm-up, and warm-down.
    \end{tablenotes}
\vspace{-0.5cm}
\end{table}

Notably, signature algorithm speed is only one of the factors impacting performance in schemes with larger key and signature sizes. For Hyperledger Fabric, the signature and public key size have an added performance cost even for fast signature algorithms, because the signatures and keys must be hashed in each block. In fact, while the Falcon-512 benchmark spent more absolute time in \texttt{Sign()} than either Dilithium-2 or qTesla-p-I, the latter two had lower throughput, with notable time spent hashing. Moreover, hashing time is not the only effect of large key and signature sizes; it is only the easiest to measure. The CPU profiles also showed increases in memcpy, proto marshalling, and memory garbage collection during the PQFabric benchmarks. There are many calling functions contributing to these increases, but one natural hypothesis is that significantly increased signature, public key, and certificate sizes increased the CPU time spent on memory management. 

Overall, it seems that maintaining a small key and signature size, even sometimes at the expense of a slower post-quantum algorithm, impacts the performance of Hyperledger Fabric. Notably, prior work \cite{campbell_transitioning_2019} examined the hypothetical slowdown to Hyperledger Fabric from post-quantum cryptography, based on the benchmarks published by NIST. The paper focused on qTesla-p-I, a reasonable choice given only the performance characteristics of \texttt{Sign} and \texttt{Verify}, but our work finds qTesla-p-I to be the worst performing of the algorithms we tested. Our work demonstrates that the CPU cycle benchmarks alone do not provide enough information to anticipate the performance impact of post-quantum cryptography in the Hyperledger Fabric production system, because of the complex performance implications of signature and public key size. 

One other interesting performance difference between the signature schemes is the range of block latencies. While ECDSA, Falcon, and Dilithium variants had relatively consistent performance across blocks ($10.0 < \sigma < 16.1$), qTesla block commit latencies varied widely ($\sigma = 33.0$). We were unable to determine the cause of this variation. 

\subsection{Future Work}
In our paper, we evaluated PQFabric with a simple one-signature endorsement policy. Changing this will affect block size by requiring more signatures per block, and it will also increase the time spent signing and verifying. Testing other policies can help establish a clearer relationship between signature and public key size versus algorithmic efficiency. 

In addition, we expect future work to focus on the hashing bottleneck we have highlighted in this paper. If public keys and signatures can be more efficiently encoded into blocks or saved elsewhere, while still preserving security properties, it may allow significant speedups for post-quantum algorithms with large public keys and signatures. Alternatively, researchers may wish to tackle the expensive block hashing by replacing Fabric's SHA2 with a hash algorithm that allows for parallel hash computations such as the \emph{ParallelHash} scheme standardized along with SHA3 \cite{kelsey2016sha}.

We have published our code on Github, and offer it to the research community at large as a testbed for experimenting with these and other optimizations. Groups proposing new hybrid signature schemes or post-quantum X.509 certificate standards may also be interested in testing their work on a high-throughput permissioned blockchain like PQFabric.

\section{Conclusion}\label{sec:conclusion}
In this work, we built PQFabric, which is to our knowledge the first version of Hyperledger Fabric whose signatures are secure against both quantum and classical computing threats. Our implementation meets practical production system requirements for backwards compatibility and cryptoagility without sacrificing its security. Through our implementation, we offer insight into incompatibilities and integration challenges that production systems are likely to face in their own quantum-safe transition. We provide benchmarks showing that, depending on the post-quantum cryptographic algorithm selected, PQFabric can achieve comparable latency and throughput performance to classical-only Fabric. Finally, we offer CPU profile analysis of our software to identify the cause of latency increases, allowing us to suggest target areas for performance improvements. We believe that our work substantially contributes to the discussion on post-quantum cryptographic standards and post-quantum blockchain integration, and hope that it will be useful both to those developing standards and those seeking to implement them.

\begingroup
\sloppy
\bibliographystyle{IEEEtran}
\bibliography{IEEEabrv,pqfabric,hyperledger-oqs, hyperledger}
\endgroup

\appendix
\subsection{Live Migration} \label{sec:appendix:migration}
Large distributed systems like blockchains require special consideration when transitioning to quantum-safe cryptography because universal downtime and a synchronous update for all nodes may not be possible. Software migrations like this one require not only assurance of comparable performance post-update, but also a careful story around backwards compatibility, gradual rollouts, and rollbacks. Though we did not attempt a live migration of a classical-crypto blockchain, our PQFabric implementation allows for a live migration with the following steps:
\begin{enumerate}
    \item Slow rollout of PQFabric software to all core blockchain nodes (orderers, peers, and endorsers). Clients do not yet need to upgrade. PQFabric is backwards-compatible, so nodes will continue signing classically until their configuration changes to hybrid mode, and verifying signatures classically until the identity provided with the signature changes to a hybrid one. They will also continue to validate X.509 certificates classically. This is intended to be a no-op update and, until step \ref{noderollout}, can be rolled back at any time. PQFabric and vanilla fabric nodes can coexist until step \ref{noderollout}.
    \item Certifying authority update. The certifying authority is given a post-quantum key and re-issues node certificates following a typical key rollover procedure. At this point, all node certificates contain an Alt-Signature-Value field, but no Alt-Subject-Public-Key-Info, because the nodes themselves do not have post-quantum keys. The nodes still do not verify the alternate signature when receiving certificates.
    \item Second node rollout for PQFabric software to read and verify alternate signature fields in X.509 certificates. This could either happen through a software update (this is the way we implemented it) or the change could be incorporated into a policy update to the transaction validation system chaincode. At this point, the nodes still do not sign their own messages with hybrid signatures, but they do verify the alternate signature field, when present, in X.509 certificates.
    \item\label{noderollout} Slow rollout of post-quantum keys to nodes, by generating a post-quantum keypair, updating the node's configuration files (including X.509 certificate), and then restarting the node. On startup, the node's MSP will read its own X.509 certificate to determine its public/private keys. It will load the post-quantum key from the certificate and begin signing with hybrid signatures. All other nodes are running PQFabric software and will verify the hybrid signature, even while they continue to sign with classical signatures.
    \item Eventual rollout of post-quantum keys and software to clients, as client integration becomes available.
\end{enumerate}

\end{document}